# Double pair breaking peak in Raman scattering spectra of triple layer cuprate Bi2223

Giulio Vincini[A], K. Tanaka[B], T. Adachi[A], L. Sobirey[A], S. Miyasaka[A], S. Tajima[A], S. Adachi[C], N. Sasaki[C], T. Watanabe[C]

[A] *Department of Physics, Osaka University, Osaka 560-0043, Japan,* [B] *Institute for Molecular Science, Okazaki, 444-8585, Japan,* [C] *Graduate School of Science and Technology, Hirosaki University, Hirosaki, 036-8561, Japan*

We report the Raman scattering measurements on the triple layer $Bi_2Sr_2Ca_2Cu_3O_{10}$ (Bi2223) crystals of four different doping levels from slightly overdoped to strongly underdoped regimes. We observed a double pair-breaking peak in the antinodal $B_{1g}$ configuration that we attribute to the two antinodal gaps opening on the outer and inner $CuO_2$-plane (OP and IP) band, respectively. The doping dependence of the pair-breaking peak energy was investigated. Considering the difference in doping level between the IP and OP, all the $B_{1g}$ pair-breaking peak energies for OP and IP were found to align on a single line as a function of doping, which is consistent with the previous results on the double and mono-layer cuprates. Within our experimental accuracy the IP and OP peaks start to appear almost at the same temperature. These findings suggest some sort of interaction between the layers. The observed gap energy is very large, not scaling with $T_c$.

One of the long standing issues in the research of cuprates superconductors is the doping dependence of the superconducting gap. While in the overdoped regime the gap size is proportional to the critical temperature $T_c$, in the underdoped regime the antinodal and nodal gaps show different doping dependence, indicating that the gap function deviates from a simple d-wave type[1–3]. With further reduction of the doping level the antinodal gap keeps on increasing although $T_c$ decreases.[4] This unusual behavior is connected to the pseudogap coexisting with superconductivity at low temperature, for which many theoretical models have been proposed.[5]

Another interesting topic of the field is the study of multilayer component, firstly and mainly because the $T_c$ of the cuprates strongly depends on the number of Cu-O plane per unit cell $n$. $T_c$ increases when $n$ increases from $n=1$ to $n=3$, where it reaches its maximum, and then decreases for $n≥4$.[6] Up to date the cause of this $T_c$ enhancement is not clear, with several possible factor being proposed, such as the tunneling of Cooper pairs between different layers[7], the increased next-nearest neighbor hopping parameter $t'$[8] and the disorder protection of the inner $Cu-O_2$ plane IP by the outer $Cu-O_2$ planes OP.[9] Additionally for $n≥3$ an interesting situation arises, namely that Cu-O planes with different doping level coexist in the same sample, where the OP is more doped than the IP due to its proximity to the charge reservoir layer CRL.[10] How these layers interact with each other, and how this affects $T_c$ is an open problem.

In the present work $Bi_2Sr_2Ca_2Cu_3O_{10}$ Bi2223, the triple layer ($n=3$) member of the BSCCO family ($Bi_2Sr_2Ca_{n-1}Cu_nO_{2n+4}$), is examined. Bi2223 features the highest $T_c$ ($T_{c,max}$ =110K) in the BSCCO family. The OP and IP are chemically inequivalent in Bi2223 because the two OP are in the pyramidal configuration with the apical oxygen and the IP is in the planar configuration with no apical oxygen. For Bi2223 we can study the electronic state at two doping levels simultaneously using one sample. Nuclear magnetic resonance (NMR) uncovered a large doping imbalance between the IP and OP of Bi2223 (p(OP)≈0.203 and p(IP)≈0.127).[11] The angle resolved photo-emission spectroscopy ARPES also revealed band splitting where one band was assigned to the IP and the other to the OP[12]. Furthermore these two bands originating from the IP and OP were found to have a different value of the superconducting gap. The multilayer effects enhancing $T_c$ is another interest to study this material.

Electronic Raman scattering ERS is a powerful technique which is bulk sensitive, momentum resolved and sensible to occupied and unoccupied states alike[13]. Using this technique, we investigated Bi2223 samples with four different doping levels, ranging from slightly overdoped to strongly underdoped. A double pair breaking peak, signature of the double superconducting gap was observed, and its doping and temperature dependence was investigated.

Bi2223 single crystals were grown by a travelling solvent floating zone method at Hirosaki[14,15], and annealed in oxygen atmosphere to control the doping levels[16]. The Raman spectra were corrected to account for the Bose-Einstein factor and are therefore proportional to the imaginary part of the Raman response function $\chi''(\omega,T)$. $B_{2g}$ and $B_{1g}$ configuration probe the nodal and antinodal region, respectively, where the incident and scattered light perpendicular to each other are along and at 45° with the Cu-O bonds respectively[13] (see inset in Fig.2(h) and (g) respectively).

Figure 1 shows the magnetic susceptibility of the samples investigated. The $T_c$ was determined from the onset temperature of the Meissner signal as 109K for the slightly overdoped and optimally doped samples, 105K for the slightly underdoped sample and 88K for the strongly underdoped sample. These will be referred as OvD109, OpD109, UnD105 and UnD88, respectively, from now on. The systematic change of $T_c$ indicates that our tuning of the oxygen content was successful. Although the $T_c$ values are nearly the same in the overdoped and optimally doped sample, the lattice parameters are different in these two sample as described later. The difference of their ERS spectra also supports the difference in doping level of these samples.

The ERS spectra for the optimally doped sample are shown in Fig.2(c,d). For the antinodal $B_{1g}$ configuration in figure 2(c) two signatures were observed when the temperature decreases from room temperature to 10K. These are: the suppression of spectral weight at low frequency below ≈600 cm$^{-1}$ and the two peaks appearing at higher frequency (≈560 cm$^{-1}$ and ≈800 cm$^{-1}$). As to the former, going from room temperature RT to 115K, we have the loss of spectral weight between 200 and 600cm$^{-1}$ that is to be attributed to the pseudogap opening[17–20]. Experimentally the pseudogap opening gives no peak in the Raman spectra, but only this kind of weak suppression of spectral weight is observed below T*. At T<$T_c$(=109K), a more dramatic suppression is observed below 500cm$^{-1}$, which is due to the superconducting gap opening.

The two peaks are associated with the Cooper pair-breaking into two Bogoliubov quasiparticles. The new observation in this work is that two pair breaking peaks are visible, and we attribute this to the double SC gap of Bi2223. Following the ARPES[12] and NMR[11] studies, the peak at lower energy is assigned to the OP and the one at higher energy is assigned to the IP. Such a double peak structure has never been reported so far in Raman spectra, and is in clear contrast with the single peak shown by the double layered $Bi_2Sr_2CaCu_2O_8$ Bi2212, [20–24] and the other double or single layered compounds[25–27]. Note that in some recent data on the triple layer Hg1223 a double pair braking peak could be visible, although it was not identified by the authors[28].

Here we note that an oxygen phonon is present at ≈590cm$^{-1}$ which may mislead us to think that the OP pair-breaking peak does not disappear above $T_C$, even though this is not the case. We have carefully measured the temperature dependence of the spectra to check whether the two peaks start to develop at different temperatures or not (see Fig.2(c)). However, within our measurement resolution, no clear difference was observed in the onset temperature for the peak development. This indicates that the two superconducting gaps open simultaneously, although the $T_c$ values are different in the IP and the OP. The energy values found here for the $B_{1g}$ peak positions are in good agreement with the ARPES data from Ref. [12].

In the $B_{2g}$ spectrum at 10K in figure 2(d) double pair–breaking peak is not visible. Instead a single, very broad peak appears. It is expected that due to the smaller values of the SC gaps in the nodal region and the originally broad feature for $B_{2g}$, the two peaks, even if they exist, overlap with each other, forming a single broad peak.

When the doping level slightly increases, the double $B_{1g}$ peaks are also clearly observed but at slightly lower energies (see Fig.2(a)). By contrast, the $B_{2g}$ peak appears at almost the same energy or a slightly lower energy as shown in Fig.2(b). A small but evident difference between the spectra of OpD109 and OvD109 proves that the doping levels of these two samples are different although the $T_c$ values are almost the same.

Next, in Fig.2(e) and (f), the spectra for the slightly underdoped sample (UnD105) are shown. In the low temperature $B_{1g}$ spectrum in Fig.2(e), the double peak feature is still visible, even though not as clearly as in the optimally and overdoped sample. Again the pseudogap opening is visible as a suppression of spectral weight between 200 and 500cm$^{-1}$, going from RT to 115K. In the $B_{2g}$ configuration in Fig.2(f), a strong but broad single peak is visible at 10K.

Figures 2(g) and (h) show the spectra for the strongly underdoped sample (UnD88). Here in the $B_{1g}$ configuration in Fig.2(g) no pair breaking peak seems visible. This suppression of the $B_{1g}$ Raman peak in the underdoped region is consistent with the previous reports for Bi2212[21,22,24,25]. It can be explained with the confinement of Cooper pairs in the antinodal region with underdoping[29] and is consistent with the tunnelling[30,31] and the ARPES data[32–34]. Here the pseudogap opening is clearly visible when the temperature decreases. Contrary to $B_{1g}$, the pair-breaking peak in nodal $B_{2g}$ configuration is clear and intense, as it can be seen in Fig.2(h).

To better view the redistribution of spectral weight due to superconductivity, we subtract the spectra just above $T_c$ from the 10K spectra. This is demonstrated in Fig.3(a) and (b) for the $B_{1g}$ and $B_{2g}$ configuration, respectively. In Fig.3(a) for the $B_{1g}$ spectra the double peak structure can be seen for most samples. For the slightly underdoped sample the double peak structure, which was not so clear from the raw spectra in Fig.2(e), becomes evident. For the strongly underdoped sample although a peak was too weak to be seen in the raw data, it becomes visible in Fig.3(a). We attribute this to the pair-breaking peak of the OP. Considering that the IP should be more underdoped than the OP and therefore suppressed more, it is reasonable that the IP pair-breaking peak does not appear. From this figure we can extract the precise peak position indicated by the dashed lines. For the $B_{1g}$ configuration the maximum of the subtracted spectra was taken as the peak position, whereas for $B_{2g}$ configuration this approach would lead to big uncertainty due to the broad peak. Therefore for $B_{2g}$ configuration we defined the peak position as the middle point between the two frequencies where the intensity is half the maximum value. While the $B_{1g}$ peak shifts to higher energy and loses intensity with underdoping, both for IP and OP, the $B_{2g}$ peak shifts to lower energy when going from the optimal to the underdoped samples. This opposite doping dependence of the peak position in the underdoped regime is consistent with the previous reports for the double and single layer cuprates and is commonly referred as two energy scale[21,22,25,26].

In order to visualize the doping dependence of the two energy gaps, we need to estimate the doping level of the IP and OP for all the samples. Although the average doping value $p$ of the sample can be obtained from the $T_c$ value, assuming the parabolic $T_c$ dependence of $p$ [35], this is a crude approximation, especially for a triple layer compound. To account for the different doping levels of the IP and OP, an alternative approach is needed. For the optimally doped sample, the IP and OP doping levels were estimated from the NMR measurement as p(OP)=0.203 and p(IP)=0.127, respectively [11]. For the two underdoped samples, however, such data are currently not available. Therefore, as a first approximation, we assume that the amount of imbalance of the doping level of the OP and IP does not change with doping. Then, the doping levels for the IP and OP can be determined from the base values for OpD109 plus the shift of the average doping level $\Delta p$ from the optimum value (p=0.16), where $\Delta p$ can be estimated from the $T_c$ value, assuming the parabolic $T_c$-$p$ relation $T_c/T_{c,max}=1-82.6\,\Delta p^2$. Namely, $p_{UnD}$(OP or IP)= $\Delta p_{Average,UnD}+p_{OpD}$(OP or IP).

For the slightly overdoped sample, since the $T_c$ is the same as the optimum value, we cannot use this method to estimate the doping level. As an alternative way, we used the c-axis lattice parameter determined by x-ray diffraction measurement. Assuming that the c-axis lattice parameter is linearly proportional to the doping level (oxygen content), we extrapolate the c-axis-$\Delta p_{Average}$ relation, obtained from the other three samples, to the overdoped side. From the c-axis value of the OvD109 sample, we roughly estimated $\Delta p_{Average}$ ($\approx 0.0097$), and calculated each layer doping of this sample as described above. The estimated average and layer doping for the IP and OP for all the samples are summarized in Table 1.

We use these $p$ values to plot the $B_{1g}$ pair-breaking peak energy as a function of the Cu-O layer doping in Fig.4. Both IP and OP $B_{1g}$ peak energies increase with decreasing $p$. The striking result is that, when the difference in doping between the two layers is taken into account, the $B_{1g}$ peak energies of the IP and OP align on a single line, giving a unifying picture of the behavior of both layers.

The $B_{2g}$ peak energy seems to be following the superconducting dome, but it is not clear whether this is a real behavior or not. Firstly, since the double peak structure is not resolved, we cannot separate the IP and OP peak energies. It is possible that two different doping dependences for OP and IP are overlapped, giving an artifact doping dependence. Additionally the originally broad $B_{2g}$ peak gives strong uncertainty on the peak position determination, and this can be seen in the large error bars in Fig.4. Finally it must be considered that we did not examine a large doping window with our four samples, to reveal a clear dome shape as in the case of other single and double layer cuprates. For all the above reasons we believe that we cannot draw strong conclusions on the doping dependence of the $B_{2g}$ peak energy. One aspect of the $B_{2g}$ spectra that seem to show a clear doping dependence is how far on the high energy side does the peak tail extends. This seems to be maximum in the optimally doped sample, while the peak tail moves at lower frequency with decreasing or increasing doping. This is especially evident with decreasing doping, suggesting a signature of the Fermi arc shrinking.

The present result is, to our knowledge, the first doping dependent spectroscopic study on the triple layer Bi2223. The doping dependence found here is qualitatively consistent with the reports on the single and double layer cuprates [21,22,25,26]. Namely, the two energy scale behavior has been confirmed also in the triple layer compound. The unusual increase of $B_{1g}$ peak energy with underdoping does not necessarily imply the increase of d-wave gap but possibly indicates the deviation from d-wave, due to the strong effect of the pseudogap in the antinodal region of the k-space, as indicated by ARPES [36,37]. The latter idea comes from the assumption of some interaction between the pseudogap and the superconducting gap. This interaction enhances the pair-breaking energy in the antinodal region which does not contribute to the superconducting condensate, whereas it suppresses $T_c$. These opposite effects of the pseudogap on the gap energy and on $T_c$ seem to hint to the superconductivity mechanism in the high $T_c$ cuprates.

Here we also introduce another scale in the right axis of Fig.4, the peak energy PE divided by $k_B*T_{c,max}$, which should be 4.2 at the optimal doping in a d-wave BCS superconductor in the weak coupling limit, if the PE corresponds to a double of the gap energy $\Delta$. In Ref. [25], it was demonstrated that this ratio of various single and double layer compounds collapse on a universal doping dependence (see Fig.2 in Ref. [25]) that is plotted by dashed lines in Fig.4. It is clear that for both $B_{1g}$ and $B_{2g}$ the PE/$k_B*T_{c,max}$ ratios are larger in Bi2223 than the case for the single or double layered cuprates. This suggests a larger energy scale of the pair-breaking peak, compared with $T_c$ in Bi2223.

The characteristic feature of Bi2223 is the coexistence of different doping layers in a unit cell, where the lower doping CuO-layer (IP) is sandwiched by the higher doping CuO-layers (OP). Experimentally we observe only a single superconductivity transition, but not a double step transition. It means that the IP and OP are not completely independent but interact with each other. The interaction between the layers is also supported by the result in Fig.4, where, despite the chemical inequivalence between the two layers, the $B_{1g}$ peak energies of the two layers align on a single line, indicating that the gap value is only controlled by doping. The interlayer coupling or a proximity effect can enhance the lower energy gap,[38] in turn reducing the higher energy gap. This averaging of the two gaps gives the peak position alignment observed here.

The origin of the large pair-breaking energy of Bi2223 cannot be concluded in the present study. However, the interlayer coupling is one of the candidates to explain it in terms of the multilayer effect. As previously described, multiple effects have been proposed to explain the high $T_c$ of the triple layer component, one or many of these can explain the gap enhancement in the two layers. These include: protection from the blocking layer disorders,[9] appropriately high next-nearest-neighbor hopping parameter t'[8] and tunneling of Cooper pairs between the layers.[7] Another possible gap enhancement mechanism is some degree of positive interaction between superconductivity and the pseudogap. With decreasing doping the superconducting gap is enhanced by the interaction with the increasing pseudogap. Therefore, the lower doping IP has a larger gap value due to its larger pseudogap. In this case the OP gap enhancement is expected from the proximity effects.[38]

In addition to these positive effects, we need to consider a negative effect of the pseudogap on superconductivity to explain the large ratio of PE/$k_B*T_{c,max}$. Since the $T_c$ is suppressed by the pseudogap in general, it is likely that the bulk $T_c$ of Bi2223 is lowered by the pseudogapped IP. The enhanced gap energy together with the suppressed $T_c$ must result in a large ratio of PE/$k_B*T_{c,max}$ in Bi2223, which suggests the potential of this triple layer compound to exhibit superconductivity at a higher $T_c$.

In conclusion we performed Raman scattering measurements on Bi2223 crystals with four different doping levels, and successfully resolved a double pair-breaking peak structure in the $B_{1g}$ spectra which was attributed to the superconducting gaps for the IP and OP. Both of the IP and OP peaks show clear doping dependences. Taken into account the difference in doping level of the IP and OP, all the $B_{1g}$ peak energies can be plotted on a single line, while the $B_{2g}$ spectra show only a broad single pair-breaking peak. The present results are the first doping dependent spectroscopic study on the triple layer cuprates where the OP and the IP signals were resolved. They are consistent with the $B_{1g}$/$B_{2g}$ gap behavior reported for various double and single layer cuprates. The IP and OP peaks were found to appear at the same temperature within our experimental resolution, and this, together with the alignment of the IP and OP layer $B_{1g}$ peak energy on a single line, hints to an interaction between the two layers. A large energy scale of the pair-breaking peaks was found in both nodal ($B_{2g}$) and antinodal ($B_{1g}$) gaps. Even though we cannot conclude its origin from the present result, it is likely that the multilayer structure with the underdoped IP causes this large gap energy scale, $PE/k_BT_c$, through enhancement of the gap and/or suppression of $T_c$ due to the effect of the pseudogap on IP.

Acknowledgments

We thank B. Loret for the useful discussion regarding their Hg1223 data.

| Sample Name | $T_c$ (K) | $p_{AVERAGE}$ | $p$(OP) | $p$(IP) |
|---|---|---|---|---|
| OvD109 | 109 | 0.1697 | 0.213 | 0.137 |
| OpD109 | 109 | 0.16 | 0.203 | 0.127 |
| UnD105 | 105 | 0.1389 | 0.182 | 0.106 |
| UnD88 | 88 | 0.1117 | 0.155 | 0.079 |

Table 1: Summary of the samples name, $T_c$, average and layer doping for the OP and IP.

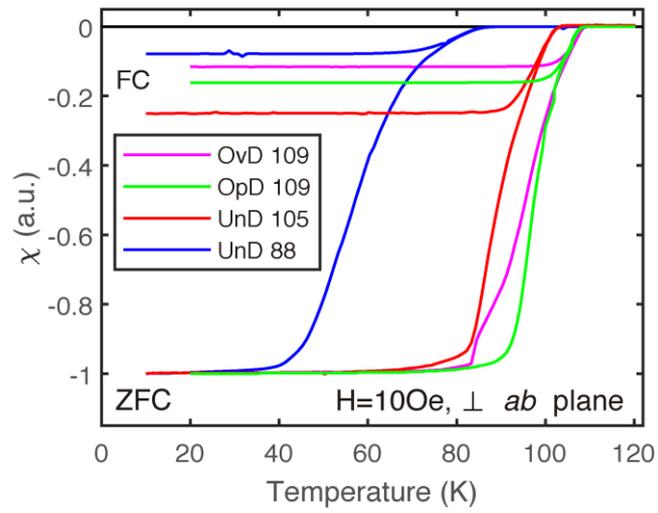

FIG. 1. (Color online) Normalized magnetic susceptibility of the four samples. The $T_c$ of all the samples were defined by the onset temperature of the Meissner signal as 109K, 109K, 105K and 88K.

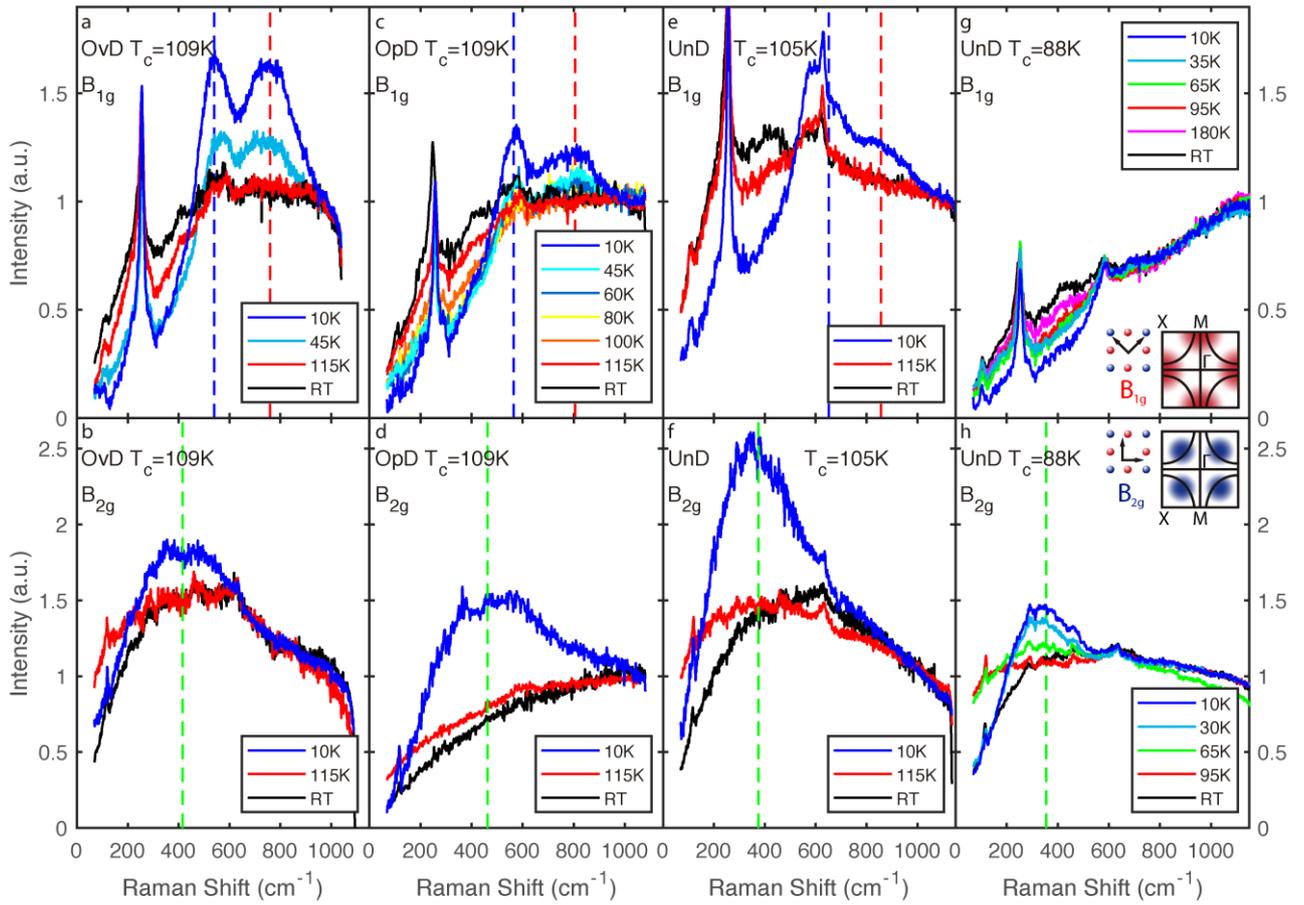

FIG.2. (Color online) $B_{1g}$ and $B_{2g}$ Raman spectra of Bi2223 for OvD109 (a,b), OpD109 (c,d), UnD105 (e,f) and UnD88 (g,h) samples. The OP and IP peak positions are indicated by blue and red dashed lines, respectively in panels a, c and e. The single peak observed in the $B_{2g}$ configuration is indicated by a green dashed line in panels b, d, f and h. The precise peak positions have been extracted by the subtracted intensity plot in Fig.3.

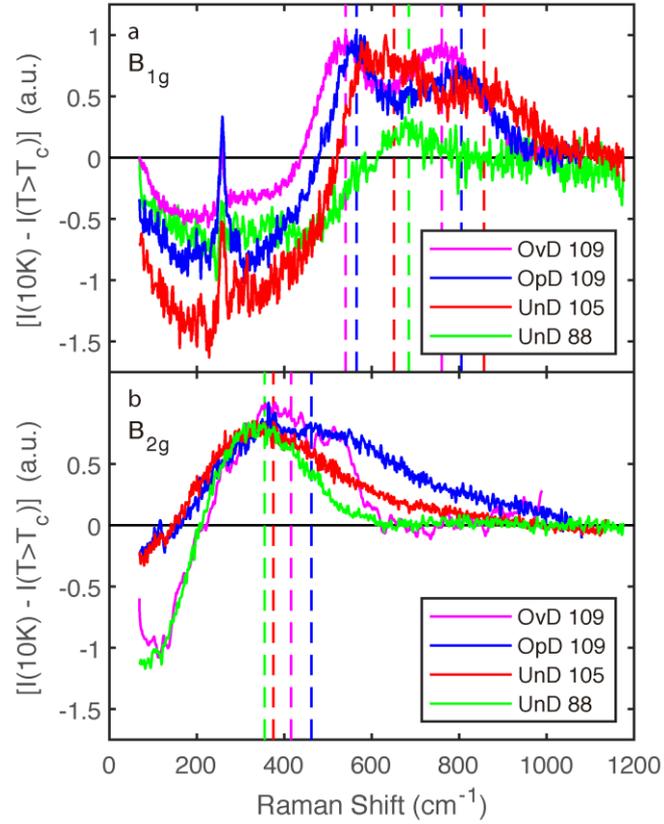

FIG. 3. (Color online) Low temperature Raman spectra of all samples after the subtraction of the spectra just above $T_C$. (a) Antinodal $B_{1g}$ high-T subtracted spectra. The double peak structure becomes clear for the OvD109 sample, OpD109 sample and the UnD105 sample. The peak of the OP becomes visible for the UnD88 sample. (b) Nodal $B_{2g}$ high-T subtracted spectra. The peak position are extracted from this figure and indicated by the dashed lines.

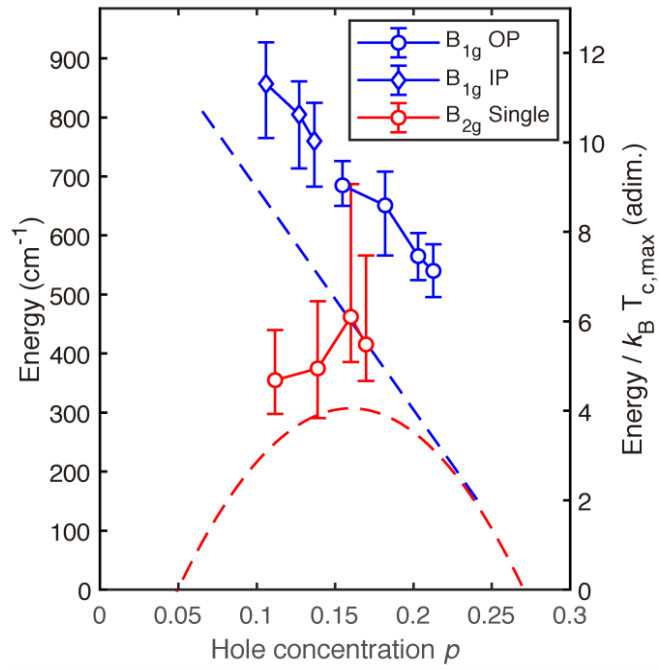

FIG. 4. (Color online) Doping dependence of the pair breaking peak energy. The Antinodal $B_{1g}$ energy is plotted using the estimated OP and IP doping. The $B_{2g}$ peak energy is plotted as a function of the estimated average doping. The dashed line and curve are the doping dependence curves taken from Fig.2 in Ref. [25] where the peak energy of single and double layer cuprates collapse when divided by $T_{c,max}$.